\documentclass[12pt]{article}
\usepackage{graphicx,amsmath,amssymb,amsfonts}
\usepackage{citesort,epsfig}

\newcommand{\PPrtNo}
{
MSU-HEP-050707 
}
\newcommand{\TITLE}
{
Parton Distributions\protect\footnote%
{Plenary talk presented at the XIII International Workshop on 
Deep Inelastic Scattering (DIS 2005), Madison WI USA, 
April 27--May 1, 2005.}
}

\newcommand{\AUTHORS} { Jon Pumplin }

\newcommand{\INST} { Department of Physics and Astronomy \\
                    Michigan State University, E. Lansing, MI 48824 }

\newcommand{\ABSTRACT}
{
{\bf Abstract.}
I present an overview of some current topics in the measurement of Parton
Distribution Functions.
}

\begin{document}



\begin{tabular}{l}
\end{tabular}
\hfill
\begin{tabular}{l}
\PPrtNo
\end{tabular}
\medskip

\begin{center}
\renewcommand{\thefootnote}{\fnsymbol{footnote}}
{
\LARGE \TITLE
}

\bigskip
{\large  \AUTHORS}

\bigskip

\INST
\end{center}

\bigskip

\ABSTRACT                 

\setlength{\parskip}{1.5 ex}
\setlength{\parindent}{2 em}



\section{Introduction}
Parton distribution functions describe the quark and gluon content of a 
hadron when the parton-parton correlations and spin structure have been 
integrated out.  
By the asymptotic freedom of QCD, the PDFs are the only 
aspect of initial-state hadron structure that is needed to calculate 
short-distance hard scattering.  Hence PDFs are a
\textit{Fundamental Measurement:}  a challenge to understand using 
methods of nonperturbative QCD; and a
\textit{Necessary Evil:}  essential input to perturbative calculations 
of signal and background at hadron colliders.

\begin{figure}[ht]
\begin{center}
  \mbox{
  \hfill
  \resizebox{0.45\textwidth}{!}{
  \includegraphics[clip=true,height=.28\textheight]{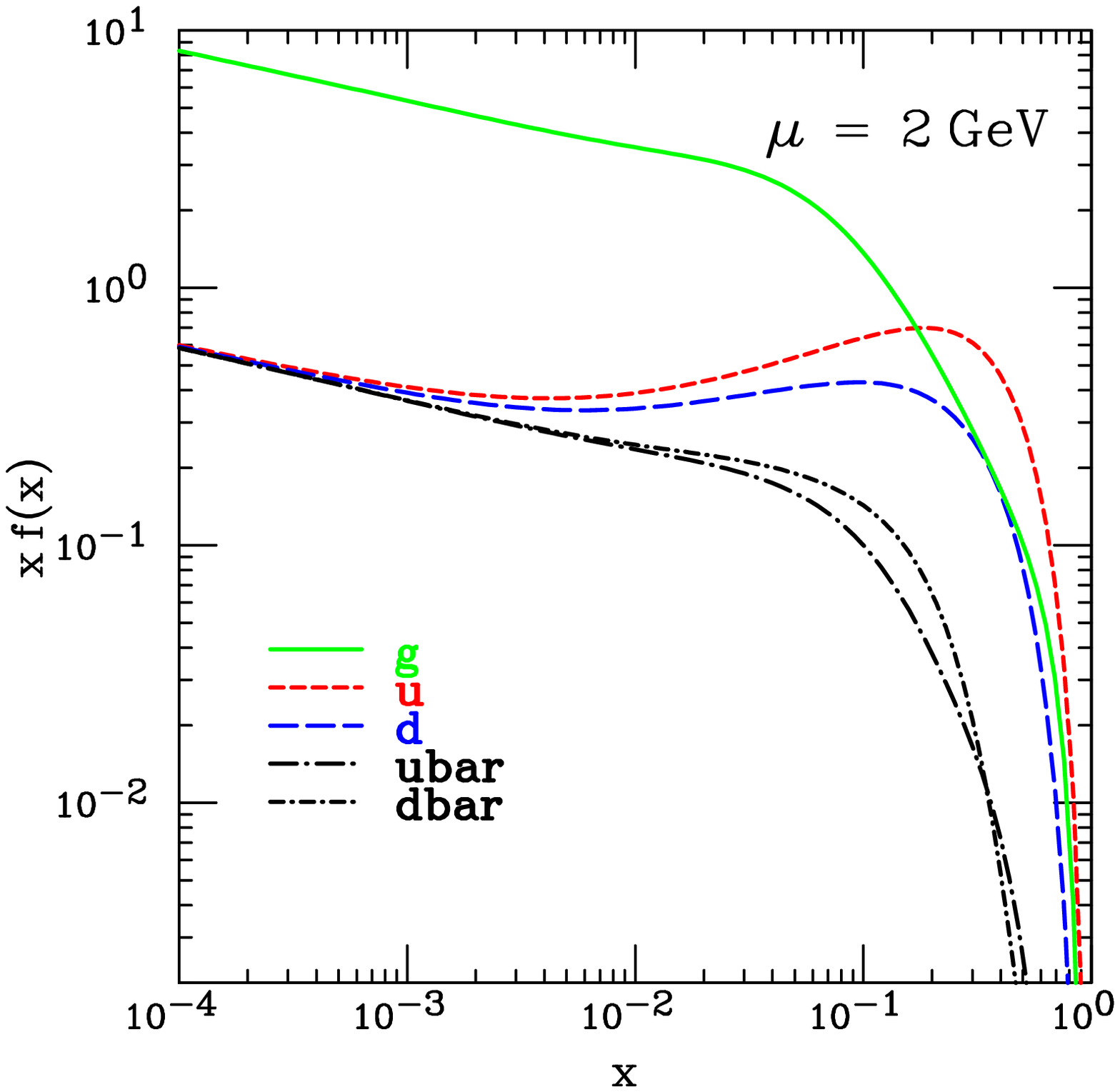}
                                     } \hfill 
  \resizebox{0.45\textwidth}{!}{
  \includegraphics[clip=true,height=.28\textheight]{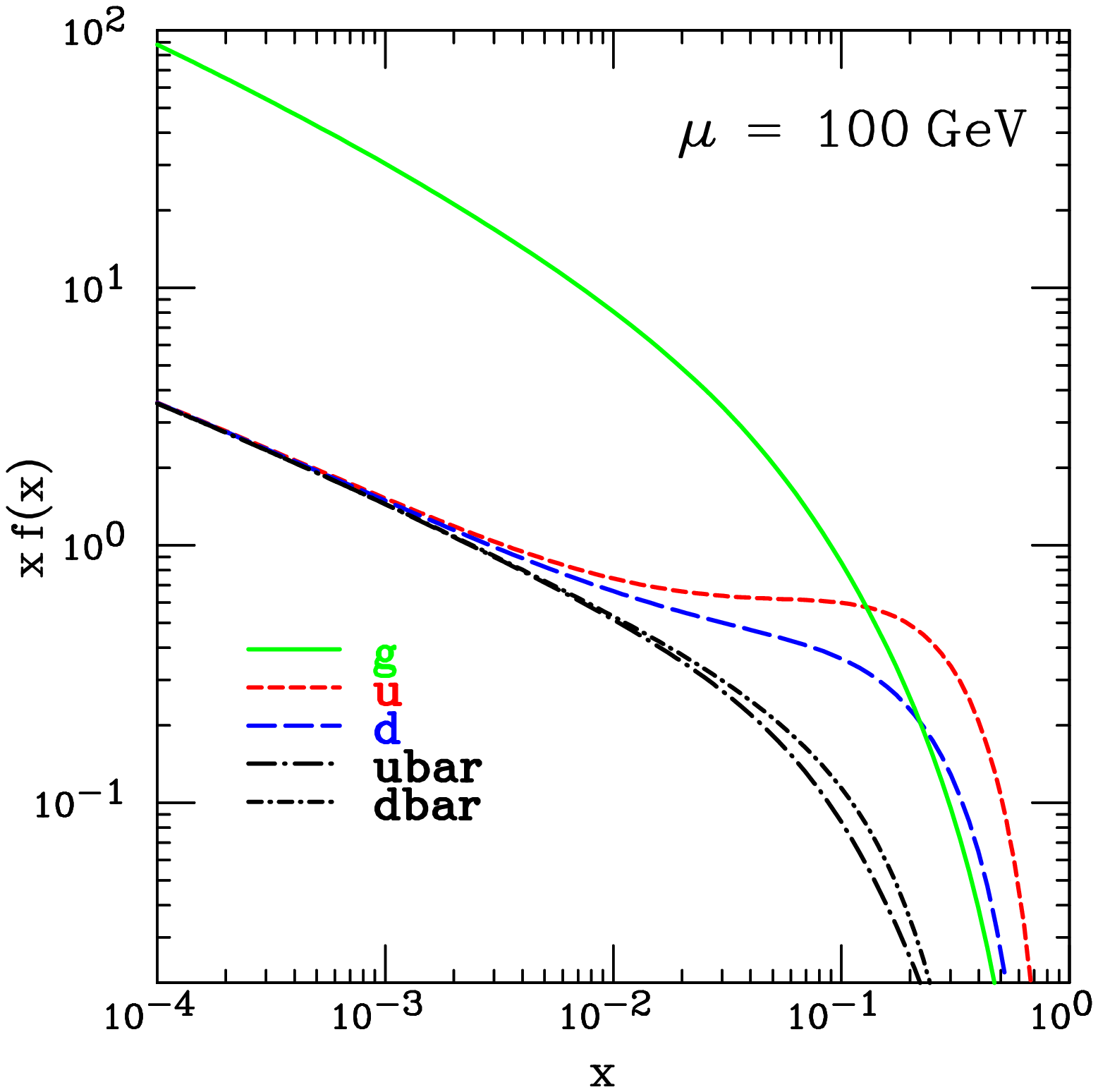}
                                     } 
  \hfill
        }
  \caption{
Overview of PDF results from CTEQ6.
  \label{fig:cteq6q}}
\end{center}
\end{figure}

The parton distributions are functions $f_a(x,\mu)$ that tell the 
probability density for a parton of flavor $a$ in a proton or other 
hadron, at momentum fraction $x$ and momentum scale $\mu$.  
An overview is shown in Fig.~\ref{fig:cteq6q} at 
$\mu=2 \, \mathrm{GeV}$ and $\mu=100 \, \mathrm{GeV}$.
Valence quarks dominate at $x \to 1$, while 
gluons dominate at small $x$, especially at large $\mu$---hence 
their vital importance for the LHC.

The PDFs are measured through a ``global analysis'' in which a large 
variety of data from many experiments that probe short distance are 
fitted simultaneously. The full paradigm consists of the following steps:
\begin{enumerate}
\item
Parameterize the $x$-dependence for each flavor at a fixed small $\mu_0$, 
using functional forms that contain ``shape parameters'' $A_1,\dots,A_N$.

\item
Compute the PDFs $f_a(x,\mu)$ at $\mu > \mu_0$ by the DGLAP equation.

\item
Compute the cross sections for DIS($e$,$\mu$,$\nu$), Drell-Yan, 
Inclusive jets, etc. by perturbation theory.

\item
Compute the ``$\chi^2$'' measure of agreement between the predictions and 
experiment:
\[
\chi^2 = \sum_i 
\left(\frac{\mathrm{Data}_i - \mathrm{Theory}_i}
{\mathrm{Error}_i}\right)^2
\]
or generalizations of that formula to include correlated experimental errors.

\item
Minimize $\chi^2$ with respect to the parameters $\{A_i\}$ 
to obtain Best Fit PDFs.

\item
Map the PDF uncertainty range as the region in $\{A_i\}$ 
space where $\chi^2$ is sufficiently close to the minimum.

\item
Make the best fit and uncertainty sets available at
http://durpdg.dur.ac.uk/HEPDATA/.  At that site, you can find 
the current CTEQ, MRST, Alekhin, H1, and ZEUS sets, along with some 
older CTEQ, MRST, and GRV sets.

\end{enumerate}

The global analysis rests on three solid theoretical pillars:
\begin{enumerate}
\item
\textit{Asymptotic Freedom} 
$\Rightarrow$ QCD interactions are weak at large scale $\mu$ (short distance), 
so a perturbative expansion in powers of $\alpha_s(\mu)$ at NLO or NNLO can be 
used to make the calculations;

\item
\textit{Factorization Theorem} 
$\Rightarrow$ PDFs are universal, i.e., the same for all processes;

\item
\textit{DGLAP evolution}
$\Rightarrow$ the dependence of $f_a(x,\mu)$ on momentum scale 
$\mu$ is perturbatively calculable, so only the dependence on 
light-cone momentum fraction $x$ for each flavor $a$ at a fixed 
small $\mu_0$ needs to be measured.
\end{enumerate}

Carrying out the global analysis brings a constant battle against 
the following challenges:
\begin{itemize}
\item
Extracting continuous functions from a finite set of measurements is 
mathematically unclean.  In particular, the analysis is carried out by 
modeling the PDFs at $\mu_0$ by smooth functions containing free parameters.
The bias that can result from the choice of these functional forms is known 
as ``parametrization dependence.''

\item
A specific type of parametrization dependence consists of simplifying 
assumptions that are made in the absence of data, such as 
strangeness symmetry ($f_s(x,\mu_0) = f_{\bar{s}}(x,\mu_0)$) which 
was assumed in most older analyses; or the condition of no intrinsic 
charm ($f_c(x,m_c) = f_{\bar{c}}(x,m_c) = 0$) which is still used.

\item
Combining data from diverse experiments is frequently made difficult by 
the presence of unknown errors. This is true even when a single discrete 
quantity like the mass or lifetime of a particle is measured; but 
the situation is worse when one is attempting to measure a large number 
of parameters (the $\{A_i\}$) from many experiments that are 
sensitive to different combinations of those parameters.

\item
The quality of the fit to data is measured---by tradition and because 
there is no obvious better alternative---by a global $\chi^2$ that is 
based on the reported experimental errors.  But unquantified experimental 
and theoretical systematic errors are found to be almost an order of 
magnitude larger, based on the level of inconsistencies observed between 
the ``pulls'' of various data sets, so the reported experimental errors 
do not really provide the ideal weighting of the data points.

\end{itemize}

\section{The landscape in \protect $x$ and $\mu$}

\begin{figure}[ht]
\begin{center}
  \includegraphics[clip=true,height=.36\textheight]{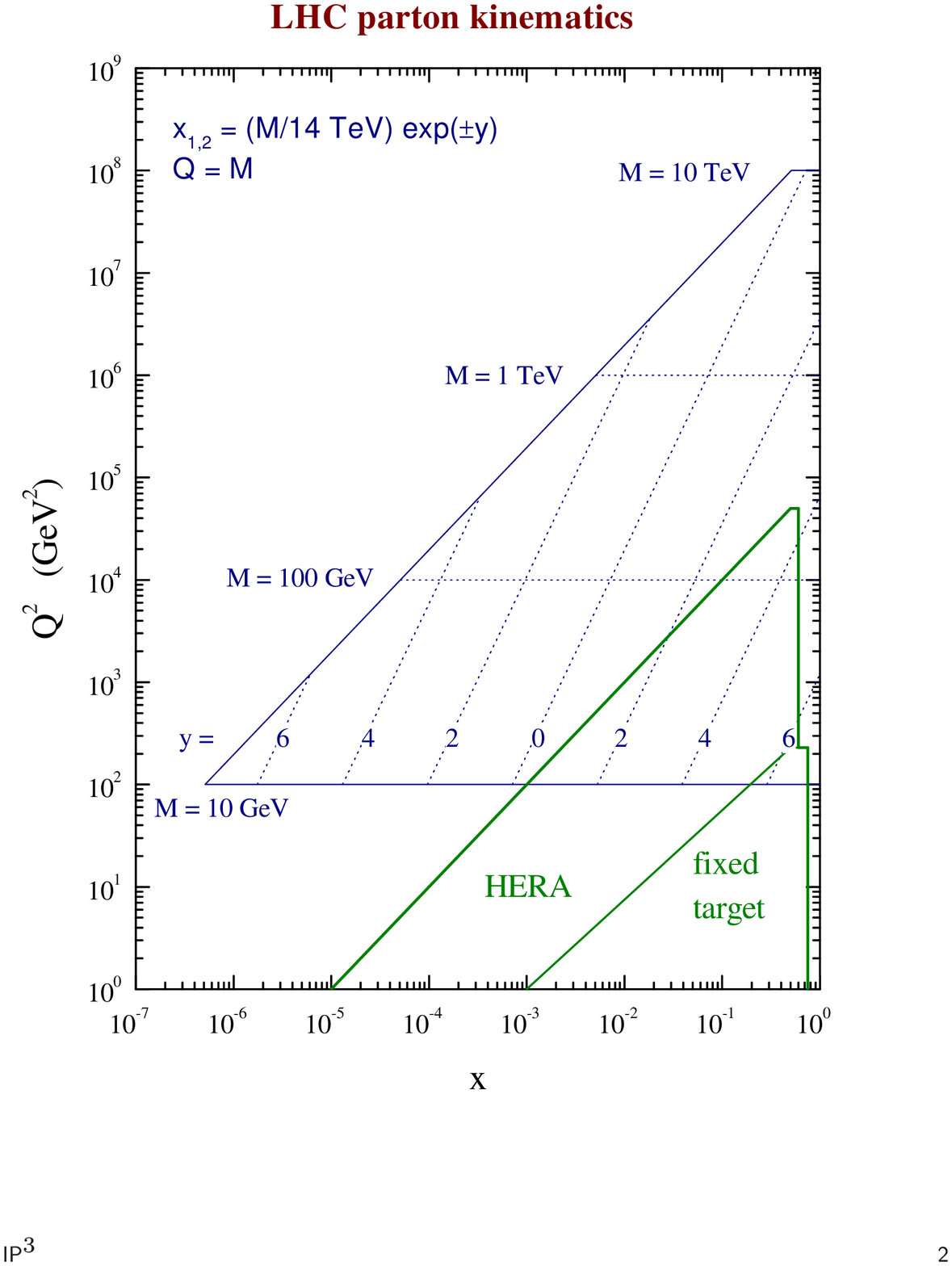}
  \caption{Kinematic map in $x$ and $Q \equiv \mu$. 
  \label{fig:KinematicMap}}
\end{center}
\end{figure}

The kinematic regions of interest in $x$ and $\mu$ are shown in 
Fig.~\ref{fig:KinematicMap} (lifted from a talk by James Stirling).  
One sees that a large range of scales in $\mu$ are connected by DGLAP 
evolution. 
The consistency or inconsistency between the different processes 
that make up the global analysis can be tested only by the global fit,
since every experiment depends differently on the PDFs.  The LHC
will dramatically extend the region of the measurements and their 
applications---especially at small $x$.

\begin{figure}[hb]
\begin{center}
  \mbox{
        \resizebox{0.333\textwidth}{!}{
        \includegraphics[clip=true,height=.14\textheight]{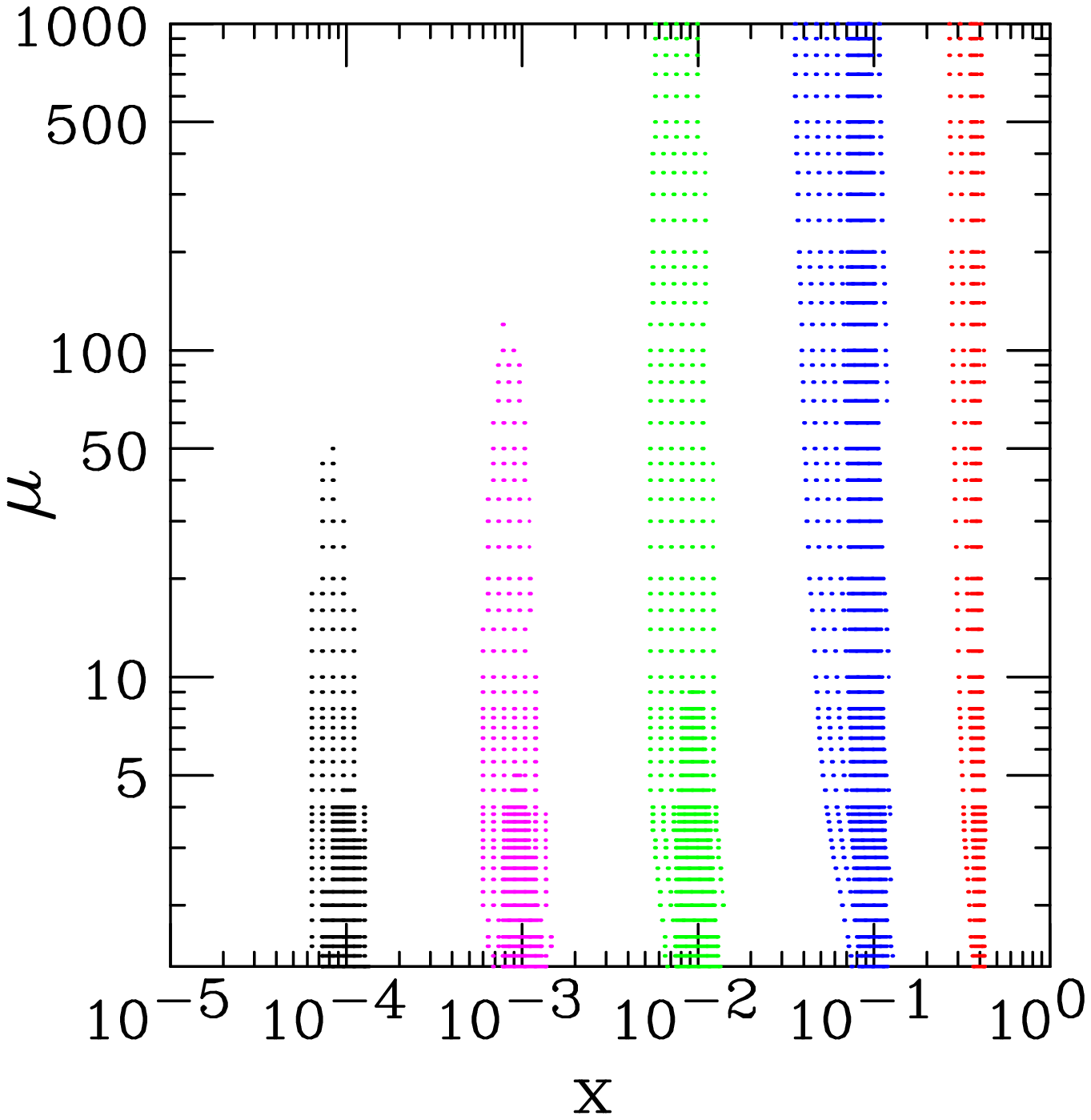}
                                     } \hfill 
        \resizebox{0.333\textwidth}{!}{
        \includegraphics[clip=true,height=.14\textheight]{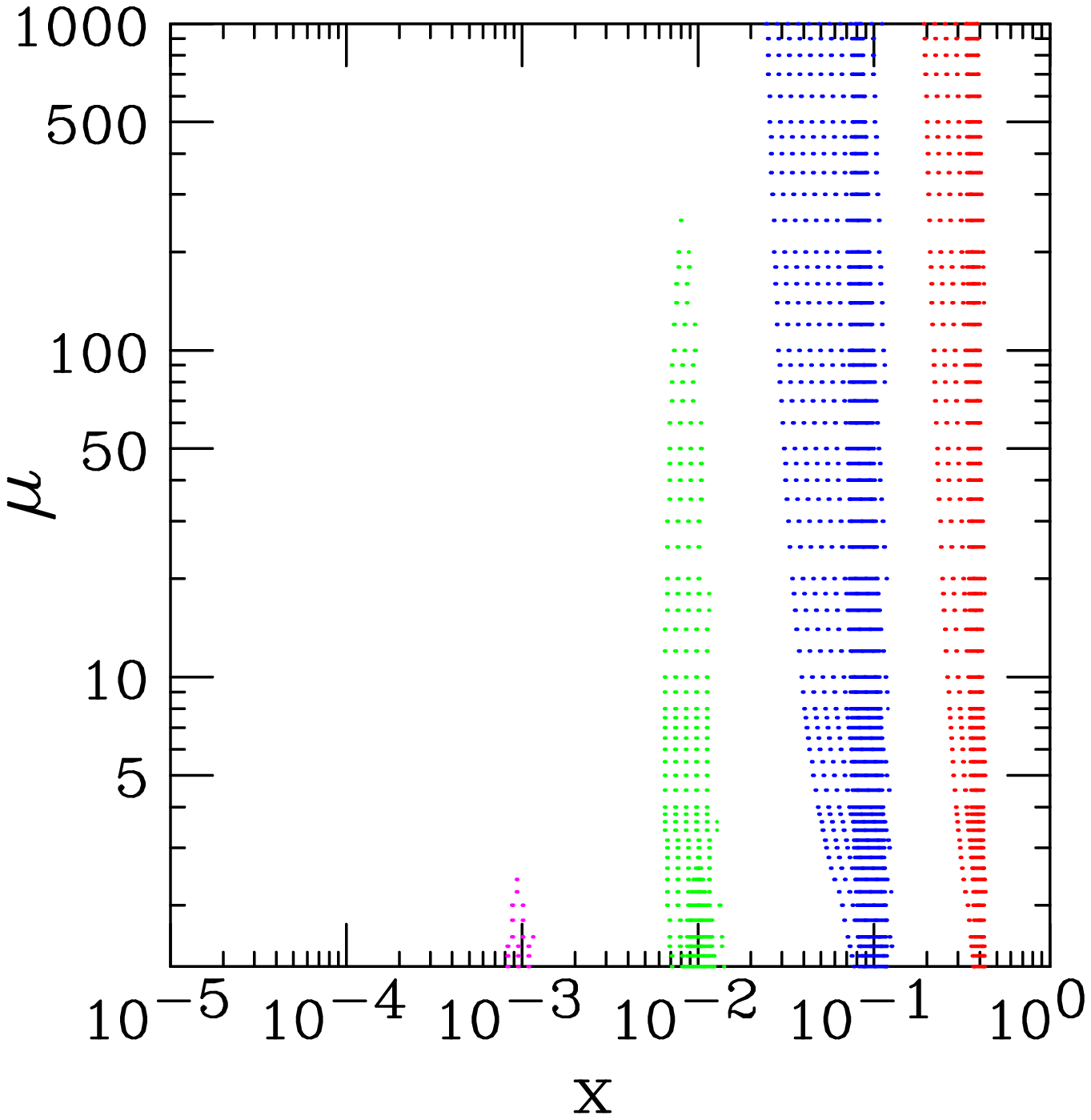}
                                     } \hfill
        \resizebox{0.333\textwidth}{!}{
        \includegraphics[clip=true,height=.14\textheight]{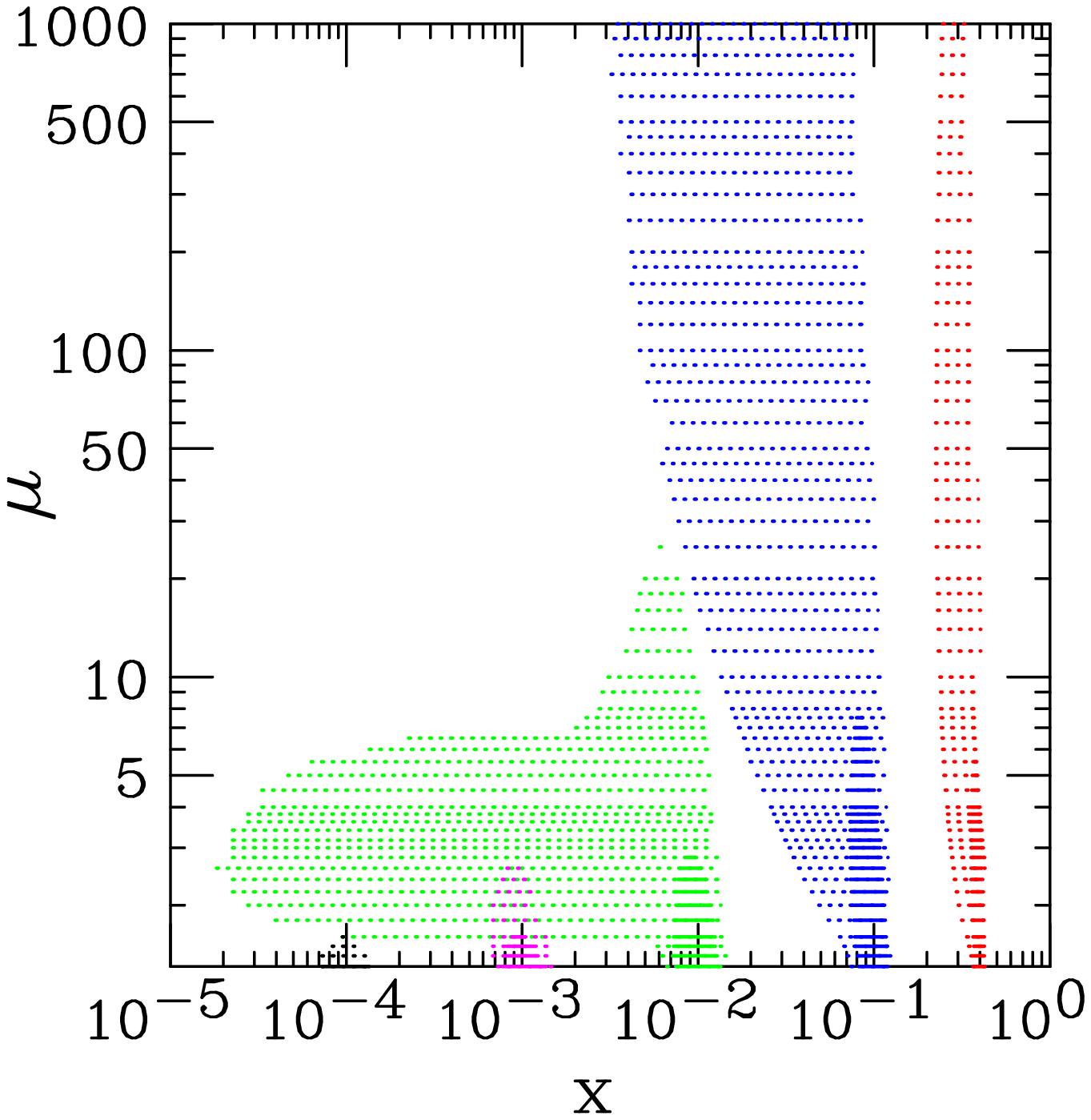}
                                     } 
        }
  \caption{
Regions of change caused by small changes in 
$\bar{u}(x) + \bar{d}(x)$ (left) or $u_v(x)$ (center) or $g(x)$ (right) 
at $\mu =1.3 \, \mathrm{GeV}$ and 
$x=10^{-4}$, $10^{-3}$, $10^{-2}$, $10^{-1}$, $0.4 \,$.
  \label{fig:probe34}}
\end{center}
\end{figure}

The DGLAP evolution in $\mu$ arises from parton branching, so the PDFs 
at a given $x$ and $\mu$ can be thought of as arising from PDFs at 
smaller $\mu$ and larger $x$.  To develop a feel for how this works
quantitatively, Fig.~\ref{fig:probe34} shows the regions where
one or more of the PDFs changes by 
$> \! 0.2 \, \%$ (solid) or $> \! 0.05 \, \%$ (dotted) 
when a $1\%$ change is made in $\bar{u} + \bar{d}$ or 
$u_v \equiv u-\bar{u}$ or $g$
at $\mu_0 = 1.3 \, \mathrm{GeV}$ in a narrow band of $x$ at various 
values of $x$.
One sees that the valence quarks are unimportant at small 
$x$---as expected---and that quark evolution is effectively at 
constant $x$, i.e., the quark distributions at a given $x$ and $\mu$ 
are mainly influenced by the quarks at $\mu_0$ at the same $x$.
The gluon at very large $x$ similarly evolves in its own world. 
But the influence of changes in the input $g(x)$ at moderate $x$ 
spreads out rapidly.\footnote{%
This results from the form of the distributions at $\mu_0$---it is 
not just a property of the DGLAP evolution.}
The small-$x$ gluon at $\mu_0 = 1.3 \, \mathrm{GeV}$ has little direct 
influence because gluons at moderate and high $\mu$ are mainly generated 
radiatively.

\section{PDF uncertainties}
A large effort has been made in recent years to intelligently consider the 
uncertainties of PDF measurements.  This is a more difficult problem than 
most other error estimates because the objects are functions, rather than 
discrete values; and because they are extracted from a complicated stew of 
experiments of different types.  Obvious sources of uncertainty are
\begin{enumerate}
\item
Experimental errors included in $\chi^2$
\item
Unknown experimental errors and theory approximations
\item
Higher-order QCD corrections $+$ Large Logs
\item
Power Law QCD corrections (``higher twist'')
\item
Parametrization dependence
\item
Nuclear corrections for data that are taken on deuterium or other nuclear 
targets.
\end{enumerate}
Essential difficulties arise from the fact that
\begin{itemize}
\item
Experiments run until systematic errors dominate, so
the remaining systematic error estimates involve guesswork.
\item
The systematic errors of the theory (e.g. power-law corrections or the 
approximations of NLO or NNLO) and their correlations are even harder to guess.
\item
Some combinations of PDFs are unconstrained, like 
$s \! - \! \bar{s}$ was before the NuTeV data.
\end{itemize}
\begin{figure}[hbt]
\begin{center}
  \includegraphics[clip=true,height=.18\textheight]{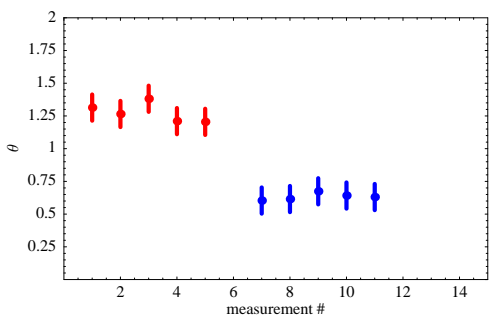}
  \caption{
Hypothetical measurements of a quantity $\theta$ by two experiments.
  \label{fig:danfig}}
\end{center}
\end{figure}
Empirically, the essence of the uncertainty problem is illustrated by the 
hypothetical Figure~\ref{fig:danfig}.
Suppose the quantity $\theta$ is measured in two different 
experiments.  What would you quote as the central value and the uncertainty?
(To play along at home, write down your answer before reading further!)
Perhaps you would expand the errors to make the uncertainty 
range cover both data sets.  Or perhaps you would expand it even more, 
using the difference between experiments as a measure of the uncertainty.
Perhaps you would also be suspicious that the spread in the points from 
each experiment is too small compared to the quoted errors.

In the global analysis, we don't get to see the conflicts so 
directly as in Figure~\ref{fig:danfig}.   But if you make a best fit 
to the points in Figure~\ref{fig:danfig} with different weights 
assigned to the two experiments, the variation of the best fit value 
with the choice of weights would map out the information that is needed. 
That approach can be used in the global fit.
In practice to quantify the uncertainties of the PDF global fit, we 
retain $\chi^2$ as the measure of fit quality, but 
vary weights of the experiments to estimate a range of 
acceptable $\Delta \chi^2$ above the minimum value, in place 
of the classical $\Delta \chi^2 = 1$.  Estimates for the current 
major global analyses are that something like 
$\Delta \chi^2 = 50 - 100$ corresponds to a $\sim 90\%$ confidence 
interval.

\section{Eigenvector uncertainty sets}
A convenient way to characterize the uncertainty of the PDFs is to create a 
collection of fits by stepping away from the minimum of $\chi^2$ along each 
eigenvector direction of the quadratic form (the Hessian matrix) that 
describes the dependence of $\chi^2$ on the fitting parameters $\{A_i\}$ in 
the neighborhood of the minimum.  This stepping is done in both directions 
along each eigenvector to allow asymmetric errors, so the 20 fitting 
parameters in CTEQ6 lead to 40 eigenvector uncertainty sets.  
The PDF uncertainty for any quantity is obtained by evaluating that quantity 
with each of the eigenvector sets and then applying a simple formula; or more 
crudely just directly from the spread in eigenvector predictions.
This method has proven so useful that generating uncertainty sets 
should be regarded as an essential part of the job for every general-purpose 
PDF determination.
In order to do this properly, CTEQ has developed an iterative 
procedure \cite{iterative}
to compute the eigenvector directions in the face of numerical instabilities 
that arise from the large dynamic range in eigenvalues of the Hessian.
(Other PDF groups have adopted the eigenvector method as well, but they 
avoid the numerical difficulties by keeping substantially fewer free 
fitting parameters, e.g. $10-15$, at a cost of greater parametrization bias.)

\begin{figure}[hbt]
\begin{center}
  \mbox{
        \resizebox{0.50\textwidth}{!}{
        \includegraphics[clip=true,height=.25\textheight]{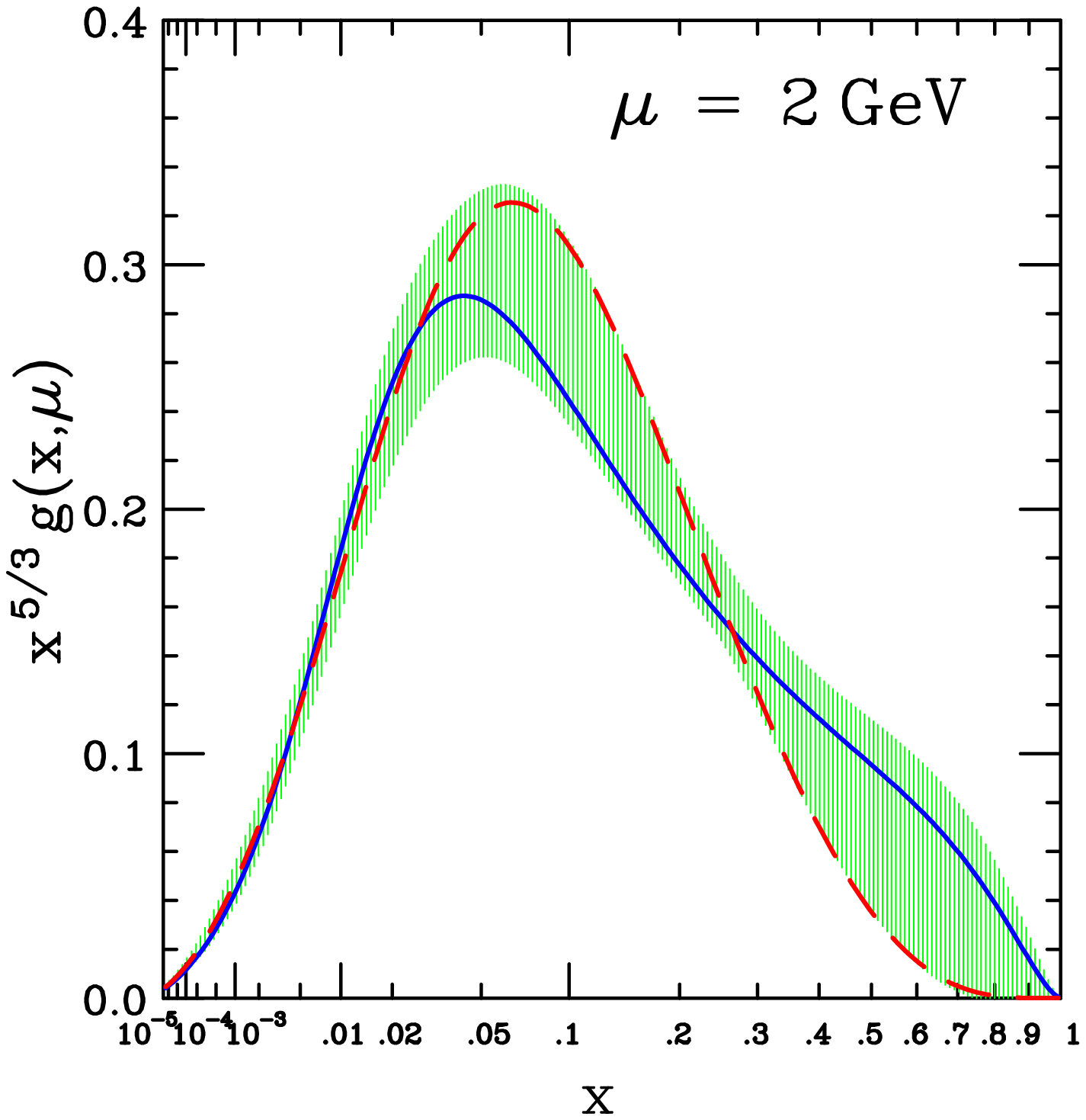}
                                     } \hfill 
        \resizebox{0.50\textwidth}{!}{
        \includegraphics[clip=true,height=.25\textheight]{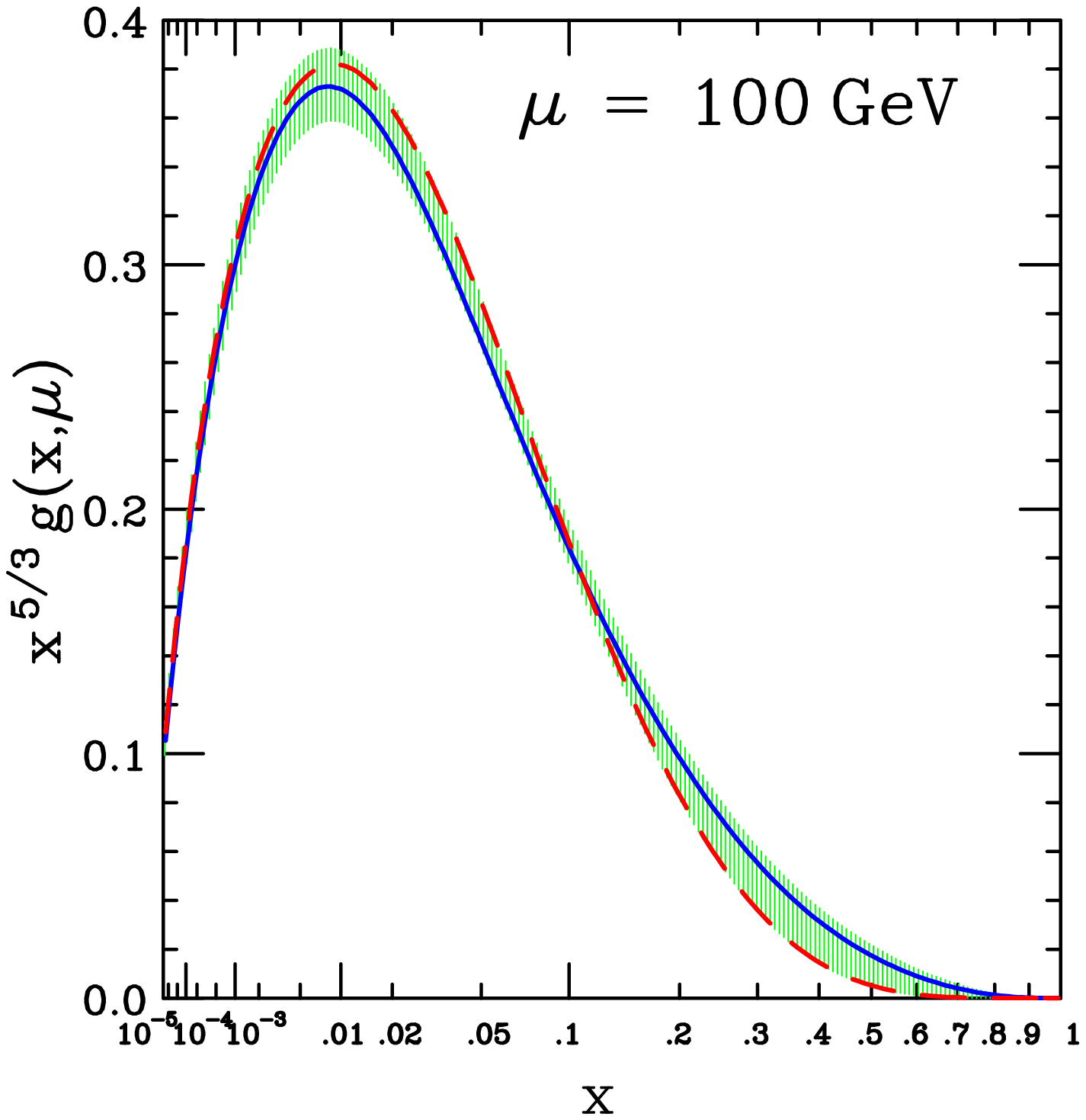}
                                     } 
        }
  \caption{
Gluon uncertainty at $\mu = 2 \, \mathrm{GeV}$ and $\mu = 100 \, \mathrm{GeV}$
from CTEQ6, plotted vs. $x^{1/3}$.  Solid (dashed) curve has extra weight for 
D\O{} (CDF) jet data.
  \label{fig:gluonAPS}}
\end{center}
\end{figure}

The uncertainty of the gluon distribution at 
$\mu = 2 \, \mathrm{GeV}$, as calculated by the eigenvector method, is shown 
in Fig.~\ref{fig:gluonAPS}.\footnote{%
The envelope of these uncertainties is not itself an allowed solution,
because the area under the curve is equal to the total gluon momentum, 
which is strongly constrained by DIS data.
Hence if $g(x)$ is larger than the central value at 
$x \approx 0.5$ it must be smaller than the central value 
at $x \approx 0.05$. 
}
Also shown are best fits in which the data were reweighted to 
emphasize D\O{} (solid) or CDF (dashed) inclusive jet cross section 
measurements.  Note that the uncertainty estimated by the eigenvector 
method is comparable to the difference between the ``pull'' of these two 
experiments.  This shows that the eigenvector method is working correctly, 
and that the jet data are a major source of the information on the gluon 
distribution.

It is interesting that these two very similar experiments pull so 
differently, which suggests that the need to allow $\Delta \chi^2 \gg 1$
may be mainly due to unknown systematic errors in the experiments.  
(In support of that notion, we also find significant differences 
between the influences of the nominally similar H1 and ZEUS components 
of DIS data.)

The right-hand side of Fig.~\ref{fig:gluonAPS} demonstrates 
``convergent evolution:''  the fractional uncertainty of the gluon is much 
smaller at large $\mu$.

\section{PDF comparisons}
\begin{figure}[hbt]
\begin{center}
  \mbox{
        \resizebox{0.333\textwidth}{!}{
        \includegraphics[clip=true,height=.15\textheight]{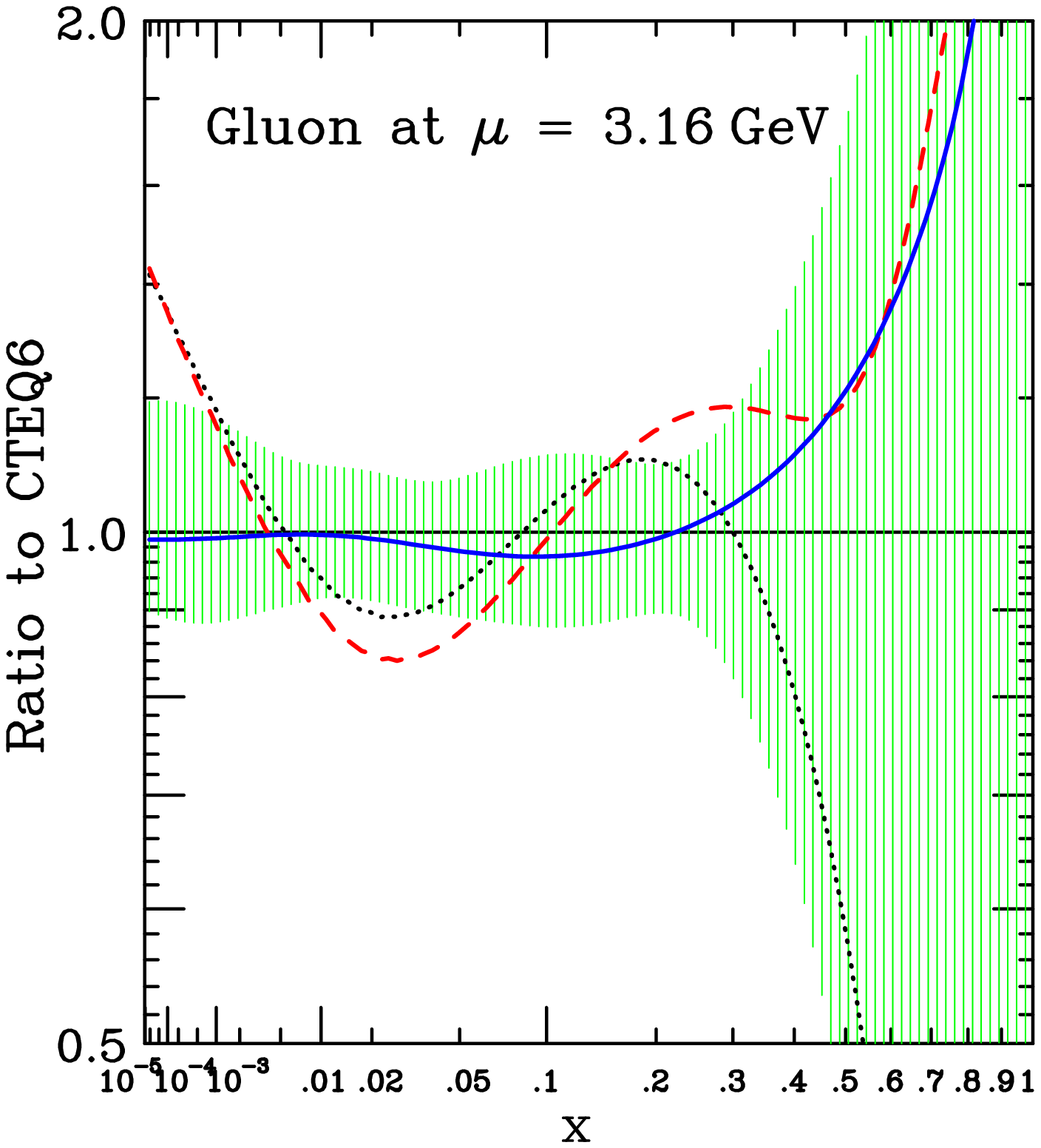}
                                     } \hfill 
        \resizebox{0.333\textwidth}{!}{
        \includegraphics[clip=true,height=.15\textheight]{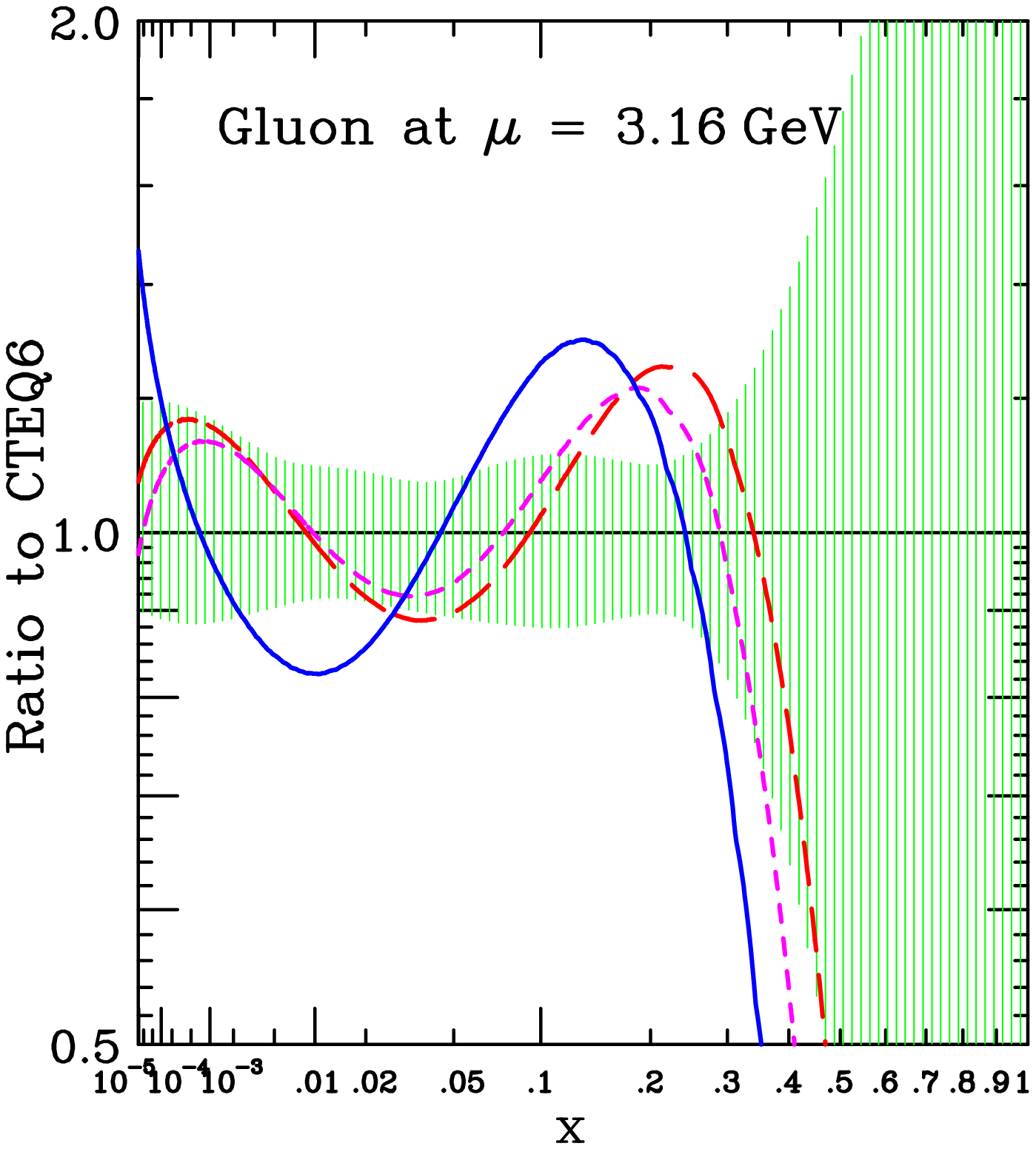}
                                     } \hfill 
        \resizebox{0.333\textwidth}{!}{
        \includegraphics[clip=true,height=.15\textheight]{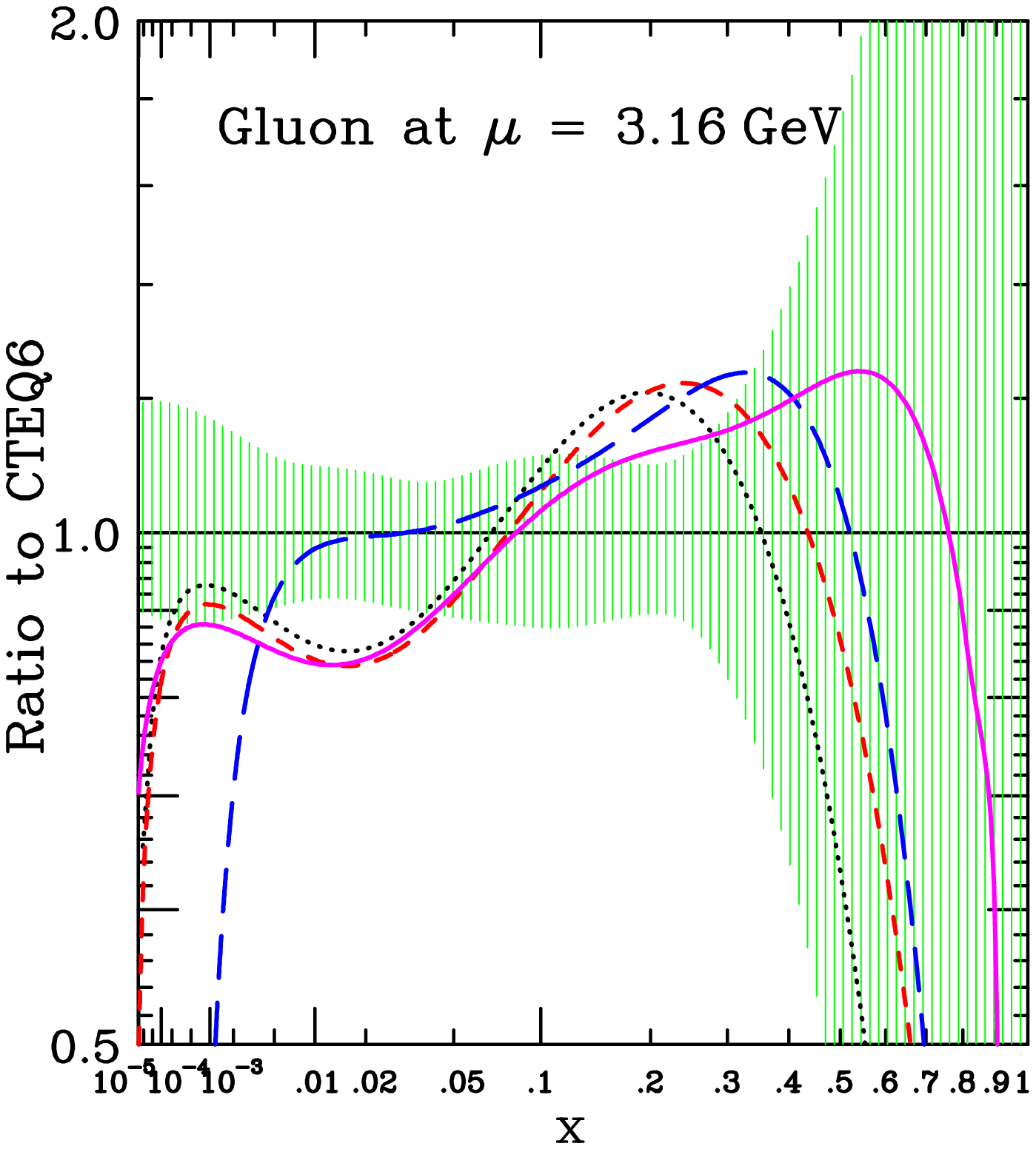}
                                     } 
        }
  \caption{
Gluon uncertainty at $\mu^2=10\,\mathrm{GeV}^2$ relative to cteq6.  
Left: cteq5 (dotted),  cteq5HJ (dashed) and cteq6.1 (solid);
Center: zeus2005zj (solid), Alekhin02NLO (long dash) and 
Alekhin02NNLO (short dash); 
Right: mrst2001 (dotted), mrst2002 (short dash), 
mrst2003 (long dash), mrst2004 (solid).
  \label{fig:GluUncQ3s}}
\end{center}
\end{figure}

The fractional uncertainty of the gluon distribution at an intermediate 
scale $\mu = 3.16 \, \mathrm{GeV}$ relative to CTEQ6 is shown in the left 
panel of Fig.~\ref{fig:GluUncQ3s}.  The dotted curve is CTEQ5, the previous 
generation of CTEQ PDFs.  The dashed curve is CTEQ5HJ, which was an early 
milestone in the PDF uncertainty business: it accounted for the seemingly 
high CDF inclusive jet cross section, relative to the QCD prediction, by 
an increased gluon distribution at large $x$ that was well within the PDF 
uncertainty range.  The solid curve is CTEQ6.1, which shows very little 
change from CTEQ6.0.

The center panel of Figure~\ref{fig:GluUncQ3s} shows a comparison with 
fits by the ZEUS group and by Alekhin.  
Since those fits are based on only a subset of the available data used in 
the CTEQ analysis, it is not surprising that they lie outside the CTEQ 
uncertainty bands.
The difference between the Alekhin NLO (long dash) and NNLO (short dash) 
is seen to be small compared to the PDF uncertainty.

The right panel of Figure~\ref{fig:GluUncQ3s} shows a comparison with fits by 
the MRST group.  
The MRST fits have progressed toward a stronger gluon at large $x$, 
which is needed to obtain good fits to the inclusive jet cross sections.
The small-$x$ behavior is sensitive to parametrization assumptions that will 
be discussed later.

It is ironic that the differences between PDF determinations by the various 
major players are comparable to the estimated uncertainty. For our original 
motivation to study the uncertainties systematically was the danger that 
comparing results from different groups might greatly underestimate the 
uncertainty, since all groups use basically the same method.  

\section{NLO and NNLO}
\begin{figure}[hbt]
\begin{center}
  \mbox{
        \resizebox{0.45\textwidth}{!}{
        \includegraphics[clip=true,height=.25\textheight]{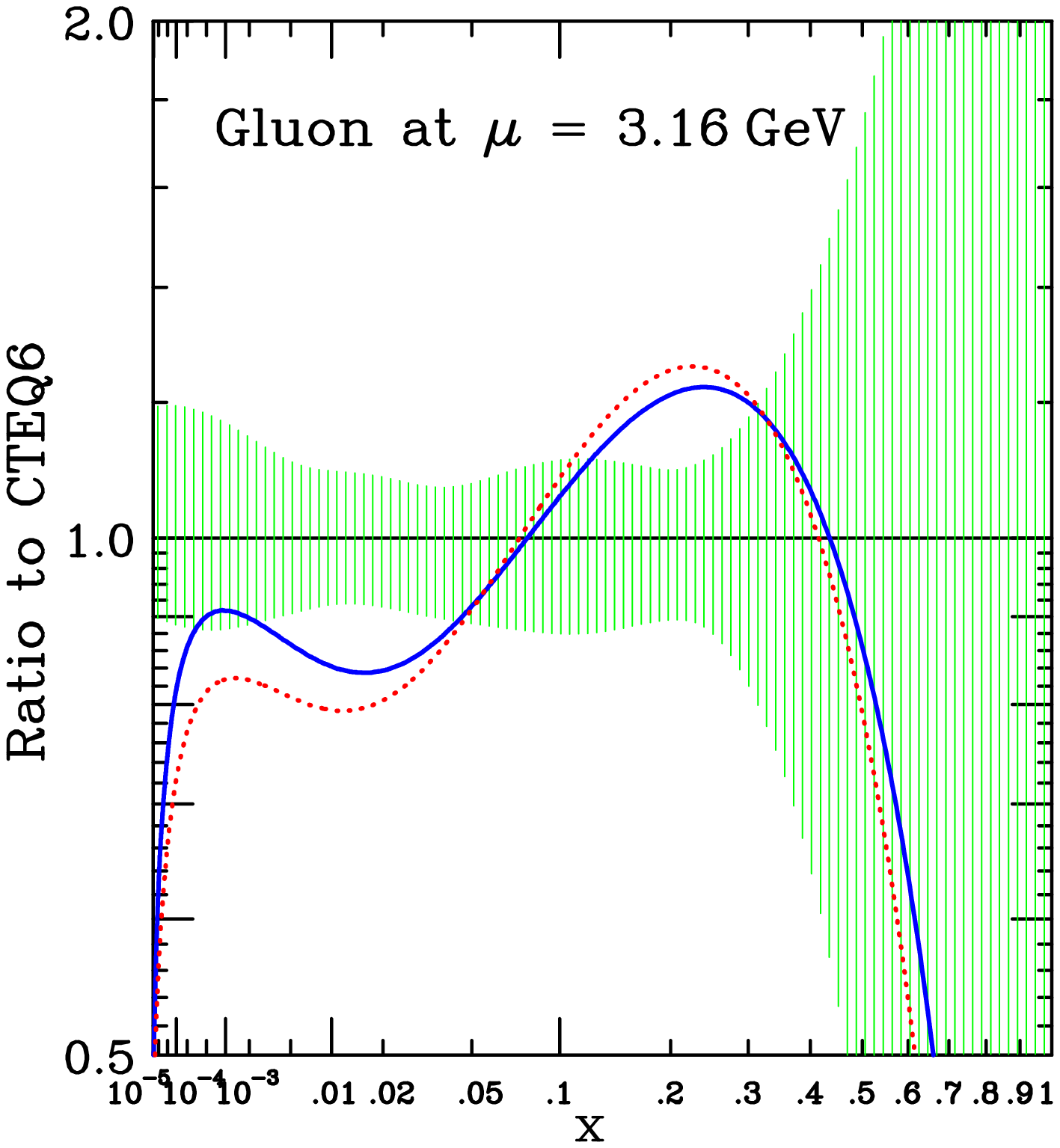}
                                     } \hfill 
        \resizebox{0.45\textwidth}{!}{
        \includegraphics[clip=true,height=.25\textheight]{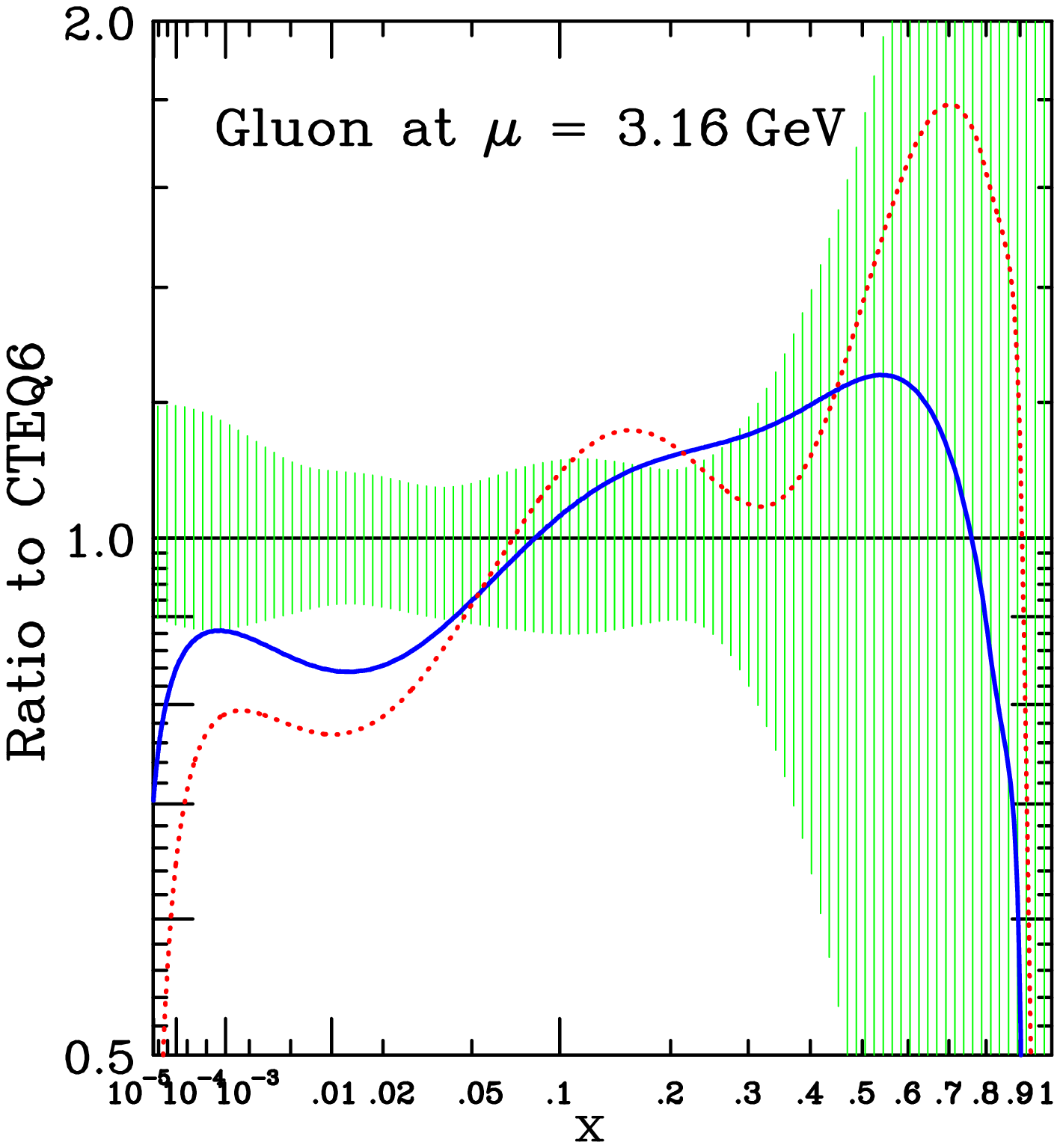}
                                     } 
        }
  \caption{Showing that NNLO effects are small for the PDFs:
Left is mrst2002 NLO (solid) and NNLO (dotted); 
Right is mrst2004 NLO (solid) and NNLO (dotted).
  \label{fig:GluUncQ3s35}}
\end{center}
\end{figure}

Figure~\ref{fig:GluUncQ3s35} shows a comparison with NLO and NNLO fits by 
the MRST group.  At present, the difference between NLO and NNLO analysis 
is small compared to the PDF uncertainty.  This is also apparent in the 
Alekhin fits shown in Fig.~\ref{fig:GluUncQ3s}.  Hence NNLO fitting, while 
obviously desirable on theoretical grounds, is not urgent.  
A reasonable goal would be to have a full set of NNLO global analysis 
tools---including jet cross sections---in place by the time LHC data 
taking begins.

\section{Negative gluons and the stability of NLO}

It has been hoped that the $W^\pm$ production cross section 
can be used as a ``Standard Candle'' to measure parton luminosity 
at the LHC.  But a challenge to the belief that $\sigma_W$ 
can be reliably predicted arose with the mrst2003c PDFs.
The `c' label refers to `conservative' cuts ($x > 0.005$, 
$Q > 3.162 \, \mathrm{GeV}$) on 
the input data set to avoid possible contamination from 
physics that is missing in the NLO analysis.  
But---as sometimes also happens in politics---this 
``conservative'' approach led to a ``radical'' outcome: 
a much smaller predicted cross section.  

\begin{figure}[hbt]
\begin{center}
  \includegraphics[clip=true,height=.30\textheight]{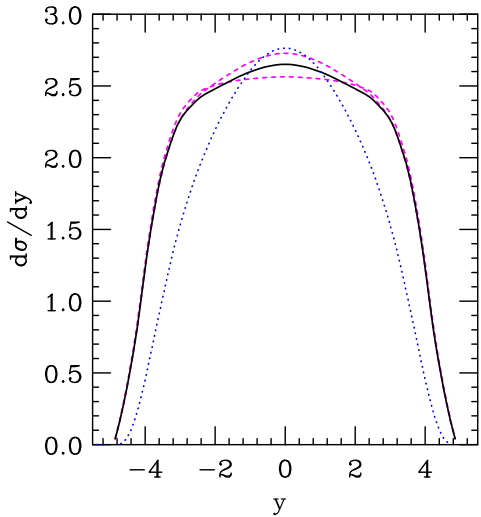}
  \caption{
Predicted rapidity distribution for $W^+ \! + \! W^-$ at the LHC:  
cteq6.1 (solid); cteq6.1 uncertainty range (dashed); mrst2003cnlo (dotted).
  \label{fig:DsigmaDy}}
\end{center}
\end{figure}

The origin of the surprising prediction is a preference of the 
MRST fit for a much smaller gluon distribution at small 
$x$, which is associated with $g(x,\mu)$ actually 
turning negative, even at a momentum scale as large as 
$\mu = m_W$.  This suppresses the predicted 
$d\sigma/dy$ for $W^+ + W^-$ at large $|y|$ as shown 
in Fig.~\ref{fig:DsigmaDy}.

In agreement with MRST, CTEQ finds that negative gluon PDFs can produce an 
acceptable fit to the data with the conservative cuts, while predicting a 
small $d\sigma/dy$ similar to that of mrst2003c.  But in disagreement with 
MRST, CTEQ finds the NLO fits to be stable with respect to variations in 
the cuts, which leaves no persuasive motivation to make the 
``conservative'' cuts.  The disagreement appears to arise from 
parametrization assumptions, since in the latest MRST NLO fit (mrst2004nlo) 
a different ansatz for the gluon distribution at $\mu_0$ has led to a 
different small-$x$ behavior.  For details, see \cite{StabilityPaper} and 
Dan Stump's talk at this conference.

%

\section{New Physics from PDF fits?}
\begin{figure}[hbt]
\begin{center}
  \includegraphics[clip=true,height=.25\textheight]{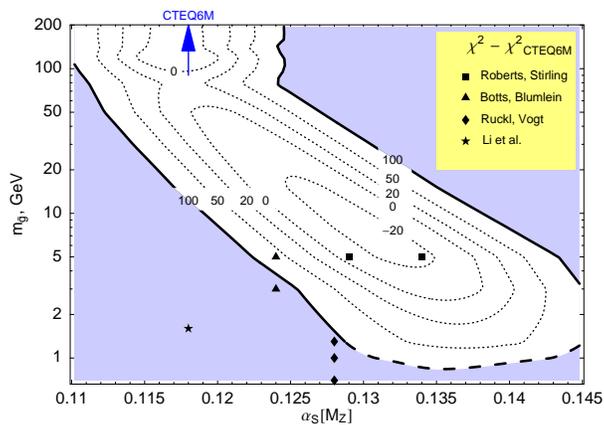}
  \caption{
Contour plot of
$\chi^2 - \chi^2_{\mathrm{CTEQ6}}$ vs. 
$M_{\tilde g}$ and $\alpha_s(M_Z)$.
  \label{fig:GluinoContours}}
\end{center}
\end{figure}

The global fit for PDFs relies on lots of Standard Model 
QCD, so the quality of the fit can be sensitive to Beyond 
Standard Model physics.  As a specific example, the 
existence of a light gluino would modify
PDF evolution and jet production \cite{gluino}.
A contour plot of $\chi^2 - \chi^2_{\mathrm{CTEQ6}}$ vs. 
$M_{\tilde g}$ and $\alpha_s(M_Z)$ is shown in 
Fig.~\ref{fig:GluinoContours}.  There is a 
valley at 
$5 \, \mathrm{GeV} < M_{\tilde g} < 20 \, \mathrm{GeV}$ with a 
depth of $\Delta \chi^2 \approx -25$, which could be regarded by 
a SUSY fanatic as a hint for a light gluino.
But a more down-to-earth interpretation is simply as a confirmation 
of the Standard Model, along with further evidence that a change 
of at least $50-100$ in $\chi^2$ is necessary to signal a 
persuasive change in the quality of the current fits. 

\section{The future}

We can expect to see steady progress in the field of parton distribution 
measurements during the final productive years of HERA and the Tevatron,
followed by dramatic developments when LHC data become available.

At the present time, there are a number of improvements underway 
from the theory side:
\begin{itemize}
\item
Improved treatment of heavy quarks.
\item
Complete NNLO calculations.
\item
Weaker input assumptions.
\end{itemize}
With regard to the input assumptions, one way to eliminate possible biases 
caused by a choice of functional 
forms for the PDfs at $\mu_0$ was discussed at this conference: to replace 
the parametrizations by neural net methods \cite{NeuralNet}.
Work is underway to open up the assumptions on strangeness: in 
previous analyses, $s+\bar{s} \propto \bar{d} + \bar{u}$ has 
been assumed.
Work is also underway to allow for possible nonradiatively generated 
(i.e., ``intrinsic'') $c, \, \bar{c}$ and $b, \, \bar{b}$.  At the same time, 
it is worthwhile to look at possibilities for making \textit{stronger} 
input assumptions, by incorporating ideas from nonperturbative models or 
lattice calculations.

There will also be many improvements in the input data set in the near future:
\begin{itemize}
\item
H1 and ZEUS are taking much more data.
\item
NuTeV data analysis at NLO is in progress.
\item
E866 final data are nearly ready. 
\item
CDF and D\O{} will have improved measurements of inclusive jets
and the lepton rapidity asymmetry from W decay.
\end{itemize}

There are some new types of measurement that could be made at HERA which 
would be useful for improving the determination of PDFs.
Most importantly, they could measure $F_L$, though it will require them to 
accept the risk of using part of their remaining running time to switch 
to a lowered proton energy.  Perhaps that risk is actually not so large, 
since the small increase in statistics to be gained by say the last half 
year of conventional running is not all that valuable.  In a more perfect 
world, it would also be very nice if HERA would measure DIS on deuterium.  

CDF and D\O{} can also make major new contributions through measurements of 
\begin{itemize}
\item
Inclusive $Z^0$ and $W^\pm$ .
\item
Inclusive jet with $c$- or $b$-tag.
\item
$\gamma/Z^0/W^\pm$ + jet with $c$- or $b$-tag.
\end{itemize}
Some data has already been reported for $Z^0$+bjet (D\O{}) \cite{Zbjet} 
and $\gamma$+bjet (CDF) \cite{Gammabjet}.
For more information on these and other possibilities, see the 
proceedings of the HeraLHC \cite{HeraLHC} and TeV4LHC \cite{TeV4LHC} 
workshops.

Principal efforts in the application of parton distributions in the 
near term will be to the difficult question of systematic errors of the 
$W$ mass measurement at the Tevatron, and to All Things LHC---the Standard 
Model and beyond.
It is also worthwhile to mention that important extensions of the PDF 
approach are going on to study (1) Spin-dependent PDFs, (2) transverse 
momentum dependent ``generalized'' PDFs, and (3) parton distributions 
of nuclei.


\paragraph*{Acknowledgements:}
 I wish to thank the organizers for an excellent conference.
 I thank Robert Thorne, Stan Brodsky, Wu-Ki Tung, Dan Stump, 
 and Joey Huston for many discussions.
 This work is supported by the National Science Foundation.



\begin{thebibliography}{9}

\bibitem{iterative}
J.~Pumplin et al.,
Phys.\ Rev.\ D {\bf 65}, 014011 (2002)
[arXiv:hep-ph/0008191].

\bibitem{StabilityPaper}
J.~Huston, et al.,
``Stability of NLO global analysis and implications for hadron collider
physics,''
[arXiv:hep-ph/0502080] to be published in JHEP.

\bibitem{gluino}
E.~L.~Berger, et al., 
Phys.\ Rev.\ D {\bf 71}, 014007 (2005)
[arXiv:hep-ph/0406143].

\bibitem{NeuralNet}
J.~Rojo et al.,
``The neural network approach to parton fitting,''
[arXiv:hep-ph/0505044] and talk at this conference.

\bibitem{Zbjet}
V.\ M.\ Abazov et al., Phys.\ Rev.\ Lett.\ {\bf 94}, 161801 (2005).

\bibitem{Gammabjet}
Reported by Amnon Harel at this conference.

\bibitem{HeraLHC}
http://www.desy.de/~heralhc/



\bibitem{TeV4LHC}
http://conferences.fnal.gov/tev4lhc/

\end{thebibliography}
\end{document}